\documentclass[a4paper,11pt]{amsart}
\usepackage{amssymb}
\begin{document}

\hyphenation{gra-vi-ta-tio-nal re-la-ti-vi-ty Gaus-sian
re-fe-ren-ce re-la-ti-ve gra-vi-ta-tion Schwarz-schild
ac-cor-dingly gra-vi-ta-tio-nal-ly re-la-ti-vi-stic pro-du-cing
de-ri-va-ti-ve ge-ne-ral ex-pli-citly des-cri-bed ma-the-ma-ti-cal
de-si-gnan-do-si coe-ren-za pro-blem gra-vi-ta-ting geo-de-sic
per-ga-mon cos-mo-lo-gi-cal gra-vity cor-res-pon-ding
de-fi-ni-tion phy-si-ka-li-schen ma-the-ma-ti-sches ge-ra-de
Sze-keres con-si-de-red tra-vel-ling ma-ni-fold re-fe-ren-ces
geo-me-tri-cal in-su-pe-rable sup-po-sedly at-tri-bu-table}

\title[Gravitational collapses and Hilbertian repulsion]
{{\bf Gravitational collapses\\and Hilbertian repulsion}}

\author[Angelo Loinger]{Angelo Loinger}
\address{A.L. -- Dipartimento di Fisica, Universit\`a di Milano, Via
Celoria, 16 - 20133 Milano (Italy)}
\author[Tiziana Marsico]{Tiziana Marsico}
\address{T.M. -- Liceo Classico ``G. Berchet'', Via della Commenda, 26 - 20122 Milano (Italy)}
\email{angelo.loinger@mi.infn.it} \email{martiz64@libero.it}

\vskip0.50cm

\begin{abstract}
If we take into account the Hilbertian gravitational repulsion,
even a ``dust'' sphere with a zero fluido-dynamical pressure
collapses to a body of a comparatively small, \emph{finite},
volume.
\end{abstract}

\maketitle


\normalsize

\vskip1.20cm \noindent \textbf{1.} -- In general relativity,
according to well known arguments, even the presence of a
fluido-dynamical pressure cannot prevent a massive body from
collapsing to a space-time singularity if a given ``point of no
return'' is passed. Now, McVittie \cite{1} and the present authors
\cite{2} have proved with \emph{exact} computations that a
spherically-symmetric gravitational collapse under the action of a
time-dependent pressure $p(t)$ ends up in a body with a finite
extent. In this paper, we shall demonstrate a very general and
very simple result: even with a zero pressure, a
spherically-symmetric collapse, because of the decisive presence
of the Hilbertian repulsive effect \cite{3}, \cite{4}, ends up in
a body of a comparatively small, \emph{finite}, volume.

\vskip0.50cm \noindent \textbf{2.} -- We start from a general
expression of the sphero-symmetrical space-time interval
$\textrm{d}s \, (c=G=1)$ \cite{5}:

\begin{equation} \label{eq:one}
\textrm{d}s^{2} = \exp\, [\nu(r,t)]\, \textrm{d}t^{2} - \exp\,
[\lambda(r,t)] \, \textrm{d}r^{2} - r^{2}
(\textrm{d}\vartheta^{2}+\sin^{2}\vartheta \textrm{d}\varphi^{2})
\quad.
\end{equation}

Then, we write the pertinent Einstein field equations with a mass
tensor $T_{j}^{k}=\varrho \, v_{j} v^{k}, (j,k=0,1,2,3)$, where
$\varrho$ is the invariant mass density of a ``dust'' sphere, and
$v^{k}$ is the four-velocity of a ``dust'' element:

\begin{equation} \label{eq:two}
R_{j}^{k} - \frac{1}{2} \, \delta_{j}^{k} R + 8\pi \varrho \,
v_{j} v^{k}=0 \quad ;
\end{equation}

(for an explicit evaluation of eqs. (\ref{eq:two}), see
\emph{e.g.} \cite{6}).

\par Eqs. (\ref{eq:two}) have as a consequence (the colon denotes a
covariant derivative):

\begin{equation} \label{eq:three}
(\varrho \, v_{j} v^{k})_{:k} = 0 \quad ,
\end{equation}

from which:

\begin{equation} \label{eq:four}
v^{k} v_{j:k} = 0 \quad :
\end{equation}

\emph{the particles of the ``dust'' sphere move according to
\textbf{geodesic} lines}.

\par For the interval $\textrm{d}s_{\textrm{ext}}$ outside the
sphere we choose the standard (Hilbert-Droste-Weyl) form of
solution \cite{3}:

\begin{equation} \label{eq:five}
\textrm{d}s^{2}_{\textrm{ext}} = \left(1-\frac{2m}{r} \right)
\textrm{d}t^{2} - \left(1-\frac{2m}{r} \right)^{-1}
\textrm{d}r^{2} - r^{2} (\textrm{d}\vartheta^{2}+\sin^{2}\vartheta
\textrm{d}\varphi^{2}) \quad,
\end{equation}

where $m$ is the mass of our sphere (if $M$ is the mass in CGS
units, we have $m=(G/c^{2})M$).

\par Let us assume for simplicity that at the beginning of the
collapse all the ``dust'' particles are at rest.

\par The surface shell of the sphere contracts itself as if the
total mass were concentrated in $r=0$. The other concentric
spherical shells contract themselves as if the mass in $r=0$ were
correspondently reduced. \emph{By virtue of the Hilbertian
repulsion} \cite{3}, \cite{4}, the particles of the superficial
shell arrive at $r=2m$ with a \emph{zero} velocity
$\textrm{d}r/\textrm{d}t$ and \emph{zero} acceleration
$\textrm{d}^{2}r/\textrm{d}t^{2}$. The particles of a generic
shell will arrive with \emph{zero} velocity and \emph{zero}
acceleration at $r=2\times$ (the relevant partial mass).
Evidently, the collapse ends when the total mass of the sphere has
filled up the spatial region $0\leq r \leq 2m$.

\par \textbf{\emph{Conclusion}}. -- We see that, on account of
Hilbert's repulsive effect, the particles of the sphere cannot go
beyond their ``points of no return'' $r=2m$, $r=2\times$ (the
relevant partial mass) -- and consequently the celebrated
arguments, which we have mentioned in sect. \textbf{1}, do not
describe the real physical situation. In the final stage the
collapsed body has a \emph{spatial extent}, it has \emph{not} been
converted into a material point.

\vskip1.50cm
\begin{center}
\noindent \small \emph{\textbf{APPENDIX}} \\ \textbf{\emph{An
instance of Hilbertian repulsion}}
\end{center} \normalsize

\vskip0.40cm \noindent For reader's convenience, we recall here
the fundamental equations of the \emph{radial} geodesics of
Schwarzschild space-time, created by a point-mass $m$. Of course,
Hilbert \cite{3} starts from the $\textrm{d}s^{2}_{\textrm{ext}}$
of our eq. (\ref{eq:five}). He puts $\alpha \equiv 2m$. Consider
Hilbert's eqs. (53) and (54):

\begin{equation} \label{eq:A1}
\frac{\textrm{d}^{2}r}{\textrm{d}t^{2}} -
\frac{3\alpha}{2r(r-\alpha)} \,
\left(\frac{\textrm{d}r}{\textrm{d}t}\right)^{2} +
\frac{\alpha\,(r-\alpha)}{2r^{3}} = 0 \quad;
 \tag{A1} \end{equation}

\begin{equation} \label{eq:A2}
\left(\frac{\textrm{d}r}{\textrm{d}t}\right)^{2} =
\left(\frac{r-\alpha}{r}\right)^{2} + A
\left(\frac{r-\alpha}{r}\right)^{3}  \quad, \tag{A2}
\end{equation}

where the constant $A$ of this first integral is negative for the
test-particles and zero for the light-rays.

\par Eq. (\ref{eq:A1}) tells us that the acceleration is negative
or positive -- \emph{i.e.}, that the gravitation acts in an
attractive or in a repulsive way -- according to the absolute
value of the velocity. Thus, when

\begin{equation} \label{eq:A3}
\left|\frac{\textrm{d}r}{\textrm{d}t}\right| < \frac{1}{\sqrt{3}}
\, \frac{r-\alpha}{r} \quad, \tag{A3}
\end{equation}

we have attraction; but when

\begin{equation} \label{eq:A4}
\left|\frac{\textrm{d}r}{\textrm{d}t}\right| > \frac{1}{\sqrt{3}}
\, \frac{r-\alpha}{r} \quad, \tag{A4}
\end{equation}

we have \emph{repulsion}.

\par Let us call $r^{*}$ the value of $r$ for which
$\textrm{d}^{2}r/\textrm{d}t^{2}=0$: attraction and repulsion
counterbalance each other. At $r=r^{*}$ the velocity has its
maximal value: $| \textrm{d}r/ \textrm{d}t| = (1/\sqrt{3}) \,
(r^{*}-\alpha)/r^{*}$.

\par For the light-rays ($A=0$), eq. (\ref{eq:A2}) gives:

\begin{equation} \label{eq:A5}
\left|\frac{\textrm{d}r}{\textrm{d}t}\right| =
\frac{r-\alpha}{r}\quad: \tag{A5}
\end{equation}

the light is \emph{repulsed everywhere}; its velocity increases
from zero at $r=\alpha$ to $1$ at $r=\infty$.

\par We see from the above equations that test-particles and
light-rays arrive at $r=\alpha$ with $\textrm{d}r/\textrm{d}t=0$
and $\textrm{d}^{2}r/\textrm{d}t^{2}=0$: \emph{the spatial
surface} $r=\alpha$ \emph{represents for them an insuperable
barrier}: a fact of paramount importance from the astrophysical
standpoint.

\par It is remarkable that attraction and repulsion are linked in a
 physical way to the radial \emph{three}-acceleration and
 the radial \emph{three}-velocity.  This means
 that the \emph{physical} evolution parameter is the ``Systemzeit'' $t$
 (von Laue) of Schwarzschild space-time,
 not the proper time of the test-particles, or the affine
 parameter of the light-rays. --


\vskip1.80cm \small
\begin{thebibliography}{9}

\bibitem{1}
G. C. Mc Vittie, \emph{Ap. J.}, \textbf{140} (1964) 401. This
author has made also interesting studies of Newtonian collapses;
see, \emph{e.g.}, \emph{A. J.}, \textbf{61} (1956) 451 -- and his
book \emph{General Relativity and Cosmology} (University of
Illinois Press, Urbana) 1965, Chapts. \textbf{6} and \textbf{7}.

\bibitem{2}
T. Marsico and A. Loinger, \emph{arXiv:physics/0612160 v1}
(December 16th, 2006).

\bibitem{3}
D. Hilbert, \emph{Mathem. Annalen}, \textbf{92} (1924) 1.

\bibitem{4}
A. Loinger and T. Marsico, \emph{arXiv:0706.3891}
$[$physics.gen-ph$]$ 16 Jul 2007; \emph{ibid.}: \emph{0710.3927}
$[$\emph{id.}$]$ 21 Oct 2007; \emph{ibid.:}\emph{0711.4997}
$[$\emph{id.}$]$ 22 Dec 2007; \emph{ibid.:}\emph{0803.0050}
$[$\emph{id.}$]$ 1 Mar 2008; \emph{ibid.:}\emph{0809.1221}
$[$\emph{id.}$]$ 7 Sept 2008; \emph{ibid.:}\emph{0904.1578}
$[$\emph{id.}$]$ 9 Apr 2009; \emph{ibid.:}\emph{0907.2895}
$[$\emph{id.}$]$ 16 Jul 2009; \emph{ibid.:}\emph{0912.1329}
$[$\emph{id.}$]$ 7 Dec 2009.


\bibitem{5}
See, \emph{e.g.}, L.P. Eisenhart, \emph{Continuous Groups of
Transformations} (Dover Publs., New York) 1961, sect.\textbf{58}.


\bibitem{6}
L. Landau et E. Lifchitz, \emph{Th\'eorie du Champ} (\'Editions
Mir, Moscou) 1966, sect.\textbf{97}.

\end{thebibliography}
\end{document}